\documentclass[aps,prl,twocolumn,notitlepage,showpacs,floatfix,superscriptaddress]{revtex4-1}
\usepackage{dcolumn}
\usepackage{tabularx}
\usepackage{bm}
\usepackage{soul}
\usepackage{amsmath,amssymb,graphicx}
\usepackage[colorlinks=true,citecolor=blue,urlcolor=blue,linkcolor=blue]{hyperref}
\usepackage{environ}

\NewEnviron{eqnsplit}{%
\begin{equation}
\begin{split}
  \BODY
\end{split}
\end{equation}
}

\newcommand{\mrm}[1]{\mathrm{#1}}

\begin{document}
%\title{Room-temperature manipulation of quantum impurity spins via dynamical modes of nanomagnets: A model study}
%\title{Room-temperature manipulation of quantum impurity spins via dynamical modes of nanomagnets}
\title{Manipulating quantum impurity spins via dynamical modes of nanomagnets}

\author{Avinash Rustagi}
\email{arustag@purdue.edu}
\affiliation{Elmore Family School of Electrical and Computer Engineering, Purdue University, West Lafayette, IN 47907, USA}

\author{Shivam Kajale}
%\email{**********}
\affiliation{Department of Electrical Engineering, Indian Institute of Technology, Bombay, Maharashtra 400076, India}
\affiliation{Media Arts and Sciences, Massachusetts Institute of Technology, Cambridge, MA 02142, USA}

\author{Pramey Upadhyaya}
\email{prameyup@purdue.edu}
\affiliation{Elmore Family School of Electrical and Computer Engineering, Purdue University, West Lafayette, IN 47907, USA}
\date{\today}
%%%%%%%%%%%%%%%%%%%%%%%%%%%%%%%%%%%%%%%%%%%%%%%%
\begin{abstract}
Quantum impurity (QI) spins offer promising information processing and sensing applications by harnessing up to room-temperature quantum coherence. Challenged by the requirement of designing local coherent drives and improving sensitivity to various signals for such applications, the search of hybrid systems coupling QI spins with matter excitations have garnered significant recent interest. We propose and theoretically study a hybrid system that couples spin-1 QI with the dynamical excitations of nanomagnets, which are controlled by mechanisms uncovered in classical spintronics. We show that in such systems the QI-spin decoherence, due to coupling to thermally excited modes of the nanomagnet, can be designed across a wide range by exploiting the control over nanomagnet's mode ellipticity and the chiral nature of the coupling between the QI spin and  nanomagnet. On the other hand, when activated electrically via voltage-induced torques, we demonstrate that QI spins can be driven coherently with large quality factors at room temperature by leveraging inherent non-linear precessional modes of the nanomagnet. Our results provide theoretical guidance for enabling unique quantum spintronic functionalities, such as local coherent driving of QI spins up to ambient conditions, and the design of nanomagnet-enhanced QI-based hybrid sensors.
\end{abstract}

\maketitle
\textit{Introduction}| Hybrid quantum systems integrate complementary advantages offered by distinct physical systems, and thereby enable exploration of a rich class of phenomena for applications in quantum information processing, communication, and sensing \cite{kurizki2015quantum,*lachance2019hybrid,*li2020hybrid,*clerk2020hybrid}. Quantum Impurity (QI) spins|paramagnetic defects in insulating hosts|coupled with the collective excitations of matter have emerged as a particularly promising candidate hybrid quantum system \cite{lee2017topicalNV}. In such platforms, the collective excitations provide resonantly enhanced control fields to drive QI-based spin qubits, as well as, transmit quantum information between them. Additionally, application of well-established quantum sensing protocols to QI spins enables hybrid nanoscale sensors of various physical signals transduced by the collective excitations.

Motivated by the challenge of enhancing coupling between QI spins and matter excitations, along with extending QI-based probing to new condensed matter systems, hybrid platforms interfacing QI spins with magnetic materials have garnered significant recent attention \cite{awschalom2021quantum}. For example, the efficient Zeeman coupling between the QI spins and magnons|the quanta of collective excitations of magnets|has been exploited and/or proposed to remotely drive \cite{labanowski2018voltage,*kikuchi2017long,*andrich2017long} and entangle \cite{trifunovic2013long,*FlebusEntangleDW,*fukami2021PRXQ} QI spins, as well as, build new sensors for probing magnetic phenomena \cite{casola2018probing} and various physical signals (magnetic \cite{trifunovic2015high}, electric \cite{solanki2022electric}, and thermal \cite{wang2018curie}).

Thus far, the QI spin-magnon hybrids have largely focused on the limit where the magnons are harbored in extended thin films. The other extreme of the collective excitations of nanomagnets is relatively less explored, which in turn offers several unique opportunities. First, the fields emanating from the dynamical modes of nanomagnet can naturally act as on-chip nanoscale microwave drives \cite{zhu2008mamr}. This could enable local driving of generic QI spins, which has thus far proved challenging \cite{wang2020electrical}. Second, the effect of dimensional confinement leads to the quantization of magnon spectrum \cite{discrete}. The corresponding suppression of Suhl instabilities \cite{andersonsuhl1955,*suhl1957theory} allows for the excitation of nonlinear modes with large precession angles \cite{mayergoyz2009nonlinear, *Ilya_Schul}, which could be leveraged for driving QI-spins in ways not possible in the extended film geometry. And last, motivated  by constructing  nanomagnet-based  low-dissipation memory, logic,  and  microwave  devices, experimental capabilities for designing, as well as, electrically coupling to the nanomagnet's modes have been uncovered in the recent past \cite{locatelli2014spin,*Zhu2012VCMA,*nozaki2012electric}. On the one hand, this offers opportunities for efficient electrical activation of the nanomagnet-generated microwave fields for coherently controlling QI spins. On the other hand, such capability can be exploited to design resonantly-enhanced QI-based hybrid electric field sensors \cite{solanki2022electric}.

\begin{figure*}[hbtp]
\centering
\includegraphics[width=0.99\textwidth]{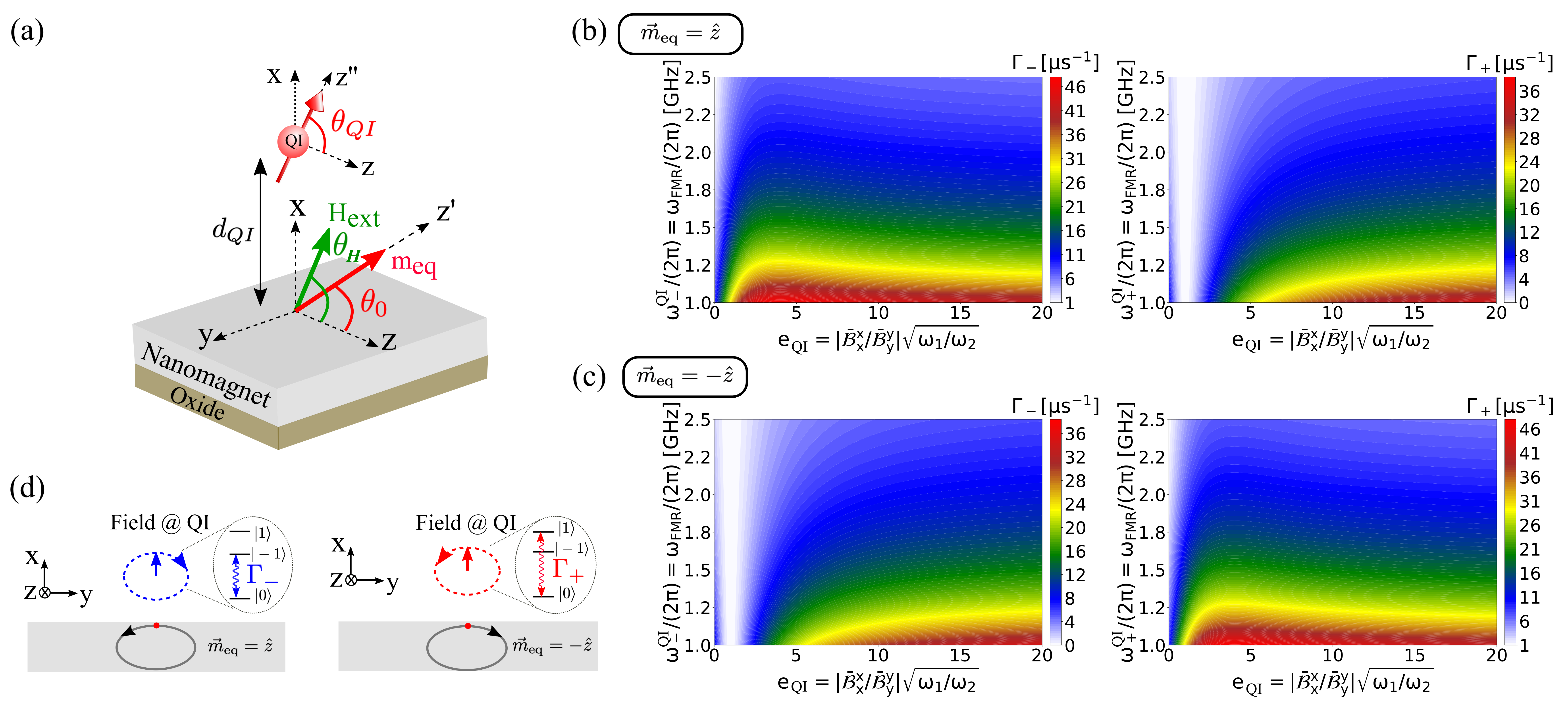}
\caption{\label{schematic_and_chiral} (a) Schematic of a QI-Nanomagnet hybrid where the external field is applied along the QI quantization axis ($\theta_{QI}=\theta_H$) and $z'$ is the direction of the magnet's equilibrium magnetization. Treating QI as an independent tunable noise spectrometer, the evaluated on-resonant QI ESR relaxation $\Gamma_-$ and $\Gamma_+$ for (b) $\vec{m}_\mrm{eq} = +\hat{z}$ and (c) $-\hat{z}$ (defined by the direction of external field) as a function of the the QI ESR transition frequency and dipolar field ellipticity $e_\mrm{QI}$, respectively. (d) The FMR precessional mode produces chiral magnetic field at the QI with opposite handedness, coupling to different QI ESR transitions when the magnet's equilibrium orientation is along $\pm z$-axis while the QI quantization axis is along $+z$-axis. Parameters are: $M_s$ = 1153 emu/cc, $H_\mrm{ext}$ = 432.26 Oe, $H_k$ = 0 Oe, $H_\perp$ = 408 Oe, $\alpha$ = 0.01, $L_y = L_z$ = 50 nm, $d_F$ = 2.2 nm, $t_{ox}$ = 2 nm.}
\end{figure*}

In this work, we theoretically study the coupling between QI spins and the dynamical modes of nanomagnets. In particular, we are motivated by the potential of enabling information processing and sensing applications that combine up to room-temperature quantum coherence of QI spins [such as those hosted within the Nitrogen vacancy (NV) centers \cite{hopper2018spin}] with the above-mentioned unique avenues offered by nanomagnets. To this end, we focus on the coupling regime where the nanomagnet's mode can be treated classically (due to high-temperature operating conditions) and are driven by electrical and/or thermal torques \cite{torque_comment}. 
Coupling to the electrically excited modes allows us to explore schemes for coherent local driving of proximal QI spins. At the same time, coupling to the thermally activated modes is central for understanding and engineering the decoherence introduced for such electrical driving, as well as, for applications using nanomagnet-induced QI spin's relaxation as a resource to sense physical signals \cite{QSensing_RMP}.

The main results of our work can be summarized as follows. First, we show that the nanomagnet-induced QI spin decoherence can be tuned across a wide range by controlling the magnon mode ellipticity and the equilibrium orientation of the nanomagnet. This tunability results from the sensitivity of QI spin and magnon coupling to the circularity and handedness (for typical spin$>$1/2 QI) of nanomagnet-generated microwave fields. Second, we find that the strength of coherent microwave fields, generated by leveraging experimentally-demonstrated electrical excitation of \textit{small-angle} dynamics in nanomagnet, can become larger than the incoherent fields due to thermal excitations. This scheme could thus be utilized to coherently drive QI spins electrically. However for such a driving scheme, the condition for maximum coupling of coherent electrical signal also coincides with maximizing decoherence, resulting in a drive with low quality factor for ambient conditions. As the third central result, we propose and demonstrate that by using electrically driven \textit{large-angle} precessional nonlinear modes of nanomagnets (such as those demonstrated experimentally for realizing voltage-induced precessional magnetization switching \cite{shiota2012LAD}), the quality factor can be improved by over two orders of magnitude. 

\textit{Hybrid Model}| Our model system consists of a QI spin in proximity to a nanomagnet (see Fig.~\ref{schematic_and_chiral}a). The Hamiltonian of the nanomagnet-QI hybrid  can be expressed as $\mathcal{H}_\mrm{hybrid} = \mathcal{H}_{QI} + \mathcal{H}_{m} + \mathcal{H}_{m-QI}$, where $\mathcal{H}_{QI}$, $\mathcal{H}_{m}$, and $\mathcal{H}_{m-QI}$ describe the Hamiltonians for QI, nanomagnet, and nanomagnet-QI interaction, respectively. Here, we consider a prototypical spin-1 QI [like Nitrogen-vacancy (NV) center] with spin triplet ground states $\vert m_s \rangle$ labeled by the projections along the QI quantization axis (at an angle $\theta_{QI}$ in the xz-plane), where $m_s = \{-1,0,1\}$. In the presence of an external field $\vec{H}_\mrm{ext}$ applied along the QI quantization axis (i.e. $\theta_{QI}=\theta_H$), the QI can be described in terms of effective two-level systems (TLSs) formed by the states ($\vert +1 \rangle, \vert 0 \rangle$) and ($\vert 0 \rangle, \vert -1 \rangle$). In this case, the QI Hamiltonian written in terms of identity $\mathcal{I}$ and Pauli $\vec{\sigma}$ matrices is $\mathcal{H}^\pm_{QI} = (\omega_\pm^{QI}/2) [\mathcal{I} \pm \sigma_z]$. Here, $\omega_\pm^{QI} = \Delta \pm \gamma H_\mrm{ext}$ are the QI-ESR frequencies with $\Delta$ and $\gamma$ labeling the zero field splitting and the gyromagnetic ratio for QI spin, respectively \cite{rustagi2020relaxometry}. 

Here we are interested in the limit where the nanomagnet-QI hybrid is operated at a temperature ($T$) much greater than that set by the nanomagnet's dynamics $\sim$ GHz, i.e. $T\gg 100$ mK. In this case, the nanomagnet can be treated classically with its Hamiltonian expressed in terms of coarse-grained magnetization vector, $\vec{M}=M_s \vec{m}$, as $H_m= \mathcal {F}(\vec{m})\mathcal {V}$. Here, $M_s$ is the saturation magnetization, $\vec{m}$ is a unit vector oriented along the magnetization, and within the single domain approximation the nanomagnet is described as a Stoner-Wohlfarth particle of volume $\mathcal{V}$ with the Hamiltonian (or equivalently the free energy) density given by $\mathcal{F} = - M_s\vec{H}_\mrm{ext} \cdot \vec{m} - M_s H_k m_z^2/2 + M_s H_\perp m_x^2/2$ \footnote{Here, for simplicity, we have focused on the regime where the nanomagnet is treated as a monodomain \cite{sun}. The formalism can be generalized to model non-uniformities by including exchange and nonuniform dipolar fields in the free energy}. The first term in $\mathcal{F}$ represents the Zeeman interaction of an external field with the nanomagnet, while the second and third terms represent in-plane and out-of-plane anisotropies arising from dipolar and/or spin-orbit interactions, which are paramaterized by the anisotropy fields $H_K$ and $H_\perp$, respectively. 

The nanomagnet-QI interaction term is captured by the Zeeman coupling between the QI-spin and the dipolar field emanating from the nanomagnet, which for the effective TLSs is of the form $\mathcal{H}^\pm_{m-QI} = \pm \gamma H^{QI}_z [\mathcal{I} \pm \sigma_z] + \gamma [H_+^{QI} \sigma_- + H_-^{QI} \sigma_+]/(2\sqrt{2})$, where $\vec{H}^{QI}$ is the dipolar field at the QI from the nanomagnet and $H_\pm^{QI} = H_x^{QI} \pm i H_y^{QI}$. The superscript `QI' denotes that the field components are evaluated in the QI-frame (where the z-axis is aligned along $z''$; see Fig.~\ref{schematic_and_chiral}a). Thermal noise-induced fluctuations of nanomagnet's magnetization produces incoherent fields at the QI and via $\mathcal{H}^\pm_{m-QI}$ decohers the QI spin. On the other hand, when the nanomagnet is driven coherently, $\mathcal{H}^\pm_{m-QI}$ allows to perform coherent manipulations of the QI spin. 

\textit{QI Relaxation}| We begin by focusing on the QI spin decoherence caused by thermally excited nanomagnet. In particular, nanomagnets host resonant modes in $\sim$ GHz frequency range. Consequently, the spectral weight of magnetic noise produced by them lies primarily in the GHz frequency range; being near to the ESR transitions of typical QI this causes population relaxation of the QI spin.  We thus focus on the (T$_1$) relaxation-type of decoherence here.  

The rates corresponding to the transitions $\vert 0 \rangle \rightarrow \vert \mp 1 \rangle$ (marked by subscript $\mp$) are given by the spectral density of the field perpendicular to the quantization axis evaluated at the ESR frequencies $\omega^{QI}_\mp$: $\Gamma_{\mp}(\omega_{\mp}^{QI}) = ({\gamma}^2/2) \int dt \, e^{i\omega_\mp^{QI} t} \big\langle H^{QI}_\pm (t) \, H^{QI}_\mp (0) \big\rangle $ \cite{rustagi2020relaxometry,Flebus2018relaxometry,Demler2019noiseMagnetometry}. Here, $\langle ... \rangle$ denotes averaging over the noise realizations. The calculation of fluctuating field in turn requires evaluation of the magnet's incoherent dynamics.
Such dynamics is governed by the stochastic Landau-Lifshitz-Gilbert (s-LLG) equation \cite{landau1980statistical,*gilbert2004phenomenological} $\dot{\vec{m}} = -\gamma \, \vec{m} \times [-\partial_{\vec{m}} (\mathcal{F}/M_s) + \vec{h}] + \alpha \, \vec{m} \times \dot{\vec{m}}$, where $\alpha$ is the Gilbert damping parameter and $\vec{h}$ is the thermal stochastic field. For typical parameters of interest to us, $h\ll -\partial_{\vec{m}} (\mathcal{F}/M_s)$; in this case s-LLG can be linearized.
Within linearized approximation, the incoherent dynamics of the magnetization components transverse to equilibrium magnetization vector (oriented at an angle $\theta_0$ and determined by the minimum of $\mathcal{F}$) is characterized by the correlations 
\begin{equation}
\label{correlator}
C^\prime_{ij}(\omega) = 2 D_{th} \sum_{\nu = \{x,y\}} S^\prime_{i\nu}(\omega) S^\prime_{j\nu}(-\omega),
\end{equation}
which is the Fourier transform of the time-domain function $C^\prime_{ij}(t-t') = \langle \delta m'_i(t) \delta m'_j(t')\rangle$ \cite{supp_ref}. Here, $\delta m'_i$ labels magnetization deviations transverse to the equilibrium (the prime superscript denotes the frame where the z-axis is aligned along the equilibrium magnetization i.e. magnet frame), $D_{th} = \alpha k_B T/(\gamma M_s L_z L_y d_F)$ \citep{brown1963thermal,kubo1970brownian} and $S^\prime_{ij}(\omega)$ denotes the susceptibility components in Fourier domain relating the magnetization deviations to the excitation field (as detailed in \cite{supp_ref}).

The dipolar field components at the QI can be related to magnetization components via the dipolar tensor $\bar{\mathcal{B}}$ i.e. $\vec{H}^{QI} = \bar{\mathcal{B}} \vec{m}'$ and $\bar{\mathcal{B}} = R_y(\theta_{QI}=\theta) \mathcal{B} R_y^T(\theta_{0})$ translates magnetization components in magnet's frame to field components in the QI frame \cite{norpoth2007straightforward,engel2005calculation,trifunovic2015high}. The QI relaxation rates can be evaluated by substituting the field components at QI into the expression for relaxation rates, giving \cite{supp_ref}
\begin{equation}
\label{RateEq}
\Gamma_\mp (\omega_\mp^{QI}) = \dfrac{\gamma^2}{2} \sum_{i,j\in \{x,y\}} \bar{\mathcal{B}}_\pm^i \bar{\mathcal{B}}_\mp^j \, C^\prime_{ij}(\omega_\mp^{QI}),
\end{equation}
where the dipolar tensor $\bar{\mathcal{B}}_\pm^k = \bar{\mathcal{B}}_x^k \pm \bar{\mathcal{B}}_y^k$ (with $k =\{x,y\}$) \cite{dipTensor_comment}. Eqs.~(\ref{correlator}) and (\ref{RateEq}) provides quantitative framework to evaluate the nanomagnet induced QI population relaxation.

\textit{Tuning QI relaxation via magnon mode's design}| Motivated by a broad range of applications in spintronics, significant progress has been made to design the energy landscape of nanomagnets. For example, by engineering interfaces and adjusting magnet's thickness, the perpendicular anisotropy ($H_\perp$) can be modulated across a wide range, even changing sign and causing spin reorientation transitions \cite{ikeda2010perpendicular}. We next apply Eq.~(\ref{RateEq}) to study and highlight the ability to tune the QI-spin's relaxation by taking advantage of this design capability offered by the nanomagnets.

In particular, we focus on the case when QI is in resonance with the thermally excited ferromagnetic resonance mode (FMR) of the nanomagnet. The FMR mode in nanomagnets can be characterized by two parameters: the resonance frequency $\omega_{\rm FMR}$ and the mode ellipticity $e \equiv \vert \delta m'_{0,x}/ \delta m'_{0,y} \vert$, where $(\delta m'_{0,x}, \delta m'_{0,y})$ is the FMR mode eigenvector expressed in terms of deviations from the equilibrium orientation. From the linearized LLG (see Eq.~(\ref{supp_LLeq}) of \cite{supp_ref}), $\omega_{\rm FMR}=\sqrt{\omega_1 \omega_2}$ and $e = \sqrt{\omega_1/\omega_2}$ where $\omega_1 = \gamma[H_\mrm{ext} \cos(\theta_H-\theta_0) + H_k \cos^2\theta_0 -H_\perp \sin^2\theta_0]$ and $\omega_2 = \gamma[H_\mrm{ext} \cos(\theta_H-\theta_0) + (H_k + H_\perp) \cos2\theta_0 ]$. In particular, the ability to tune the nanomagnet's energy landscape (for example, by tuning anisotropies) translates into controlling the resonance frequency and ellipticity of the FMR mode. In Fig.~\ref{schematic_and_chiral}b we summarize the effect of this control on the relaxation of a QI proximal to it for typical geometrical and material parameters (see the caption). Here, we have evaluated the relaxation for the case when the relevant QI ESR transition and the FMR mode are resonant with each other. Furthermore, we have scaled $e$ by the ratio of dipolar tensor components $\vert \bar{\mathcal{B}}_x^x/\bar{\mathcal{B}}_y^y \vert$ to convert the FMR mode ellipticity into the ellipticity of magnet's dipolar field at the QI (represented as $e_{QI}$ along the x-axis).

\begin{figure*}[hbtp]
\centering
\includegraphics[width=0.99\textwidth]{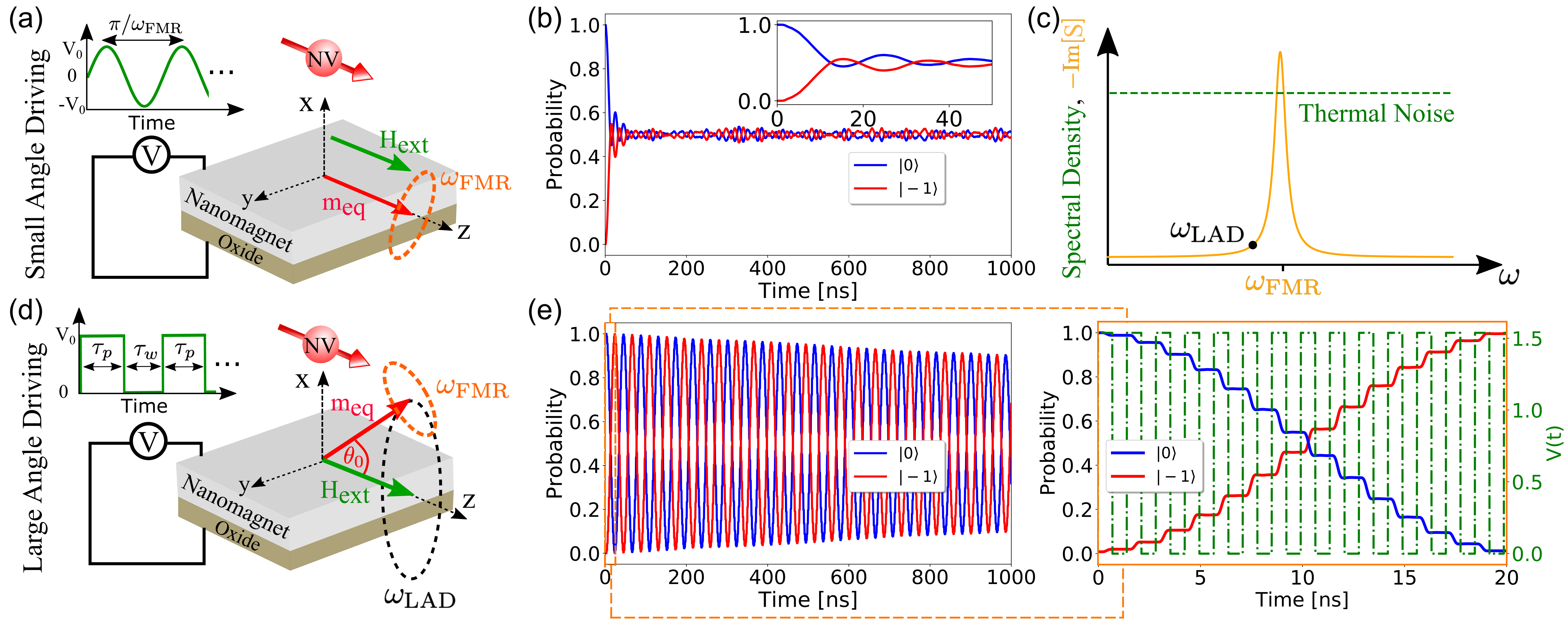}
\caption{\label{ParametricLAD} (a) Schematic of the NV QI-easy-plane nanomagnet hybrid with NV quantization axis oriented along the z-axis and the equilibrium magnetization oriented along the z-axis determined by the applied field. The magnet is driven parametrically using a sinusoidal voltage pulse $V(t) = -1.54 \sin(2\omega_\mrm{FMR} t)$ as shown. (b) NV TLS dynamics induced by the dipolar field from the driven nanomagnet at the FMR frequency in presence of thermal stochastic field [Parameters: same as in the caption of Fig.~\ref{schematic_and_chiral} and $\beta$ = 150 fJ/Vm \cite{nozaki2019recent}]. Inset shows dynamics on a shorter timescale. (c) The nanomagnet's response to perturbations is captured by the imaginary part of its susceptibility $- \text{Im}[S] \propto - \text{Im}[1/(\omega - \omega_\mrm{FMR} + i \Delta\omega_\mrm{FMR})]$ which is populated by the white noise spectral density of thermal fluctuations. (d) Schematic of the NV QI-out-of-plane easy axis nanomagnet hybrid with NV quantization axis oriented along the z-axis and the equilibrium magnetization oriented in the x-z plane at angle $\theta_0$. The magnet here is driven using a periodic step-like voltage pulse on(off) for time $\tau_p$ ($\tau_w$) as shown which induces the magnet's large angle dynamics. (e) NV TLS dynamics induced by the dipolar field from the driven nanomagnet at the LAD frequency [Parameters: $H_\perp$ = -1000 Oe, $H_\mrm{ext}$ = 502.16 Oe, and $d_F$ = 2 nm. Others remain the same]. A close up look into the TLS dynamics shows step-like Rabi dynamics.} 
\end{figure*}

First, we note that over a broad range of frequencies and ellipticities, for typical QI (like the NV center), the relaxation rate of QI is dominated by the thermal bath provided by the nanomagnet (the obtained values of relaxation rate are about 2-3 orders of magnitude larger than $\sim$ms$^{-1}$ relaxation rate of NV center in the absence of the nanomagent \cite{ariyaratne2018nanoscale}). Second, as $\omega_{\rm FMR}$ increases, this rate decreases monotonically for both ESR transitions, which can be understood as follows. With increasing resonance frequency the thermal population of the mode decreases, equivalently resulting in smaller deviations of magnetization due to the thermal noise. Thereby, the fluctuating dipolar fields at the QI due to nanomagnet are weaker for higher $\omega_{\rm FMR}$. 

Third, and more interestingly, the QI relaxation varies non-monotonically with the mode ellipticity, which shows remarkably different behavior for $\vert 0 \rangle \rightarrow \vert +1 \rangle$ and $\vert 0 \rangle \rightarrow \vert -1 \rangle$ ESR transitions. Namely, as $e_{\rm QI}$ increases from zero, $\Gamma_+$ decreases first, becomes negligibly small around $e_{\rm QI} \sim 1$, and then increases again. On the other hand, $\Gamma_-$ remains large through all values of $e_{\rm QI}$ (c.f. Fig.~\ref{schematic_and_chiral}b ). Such dependence can be understood as a consequence of chiral coupling \cite{rustagi2020relaxometry} between the QI spin and the nanomagnet. That is, for $\gamma H_{\rm{ext}} < \Delta$ the spin-1 QI have opposite built-in effective fields for the $\vert 0 \rangle \rightarrow \vert +1 \rangle$ and $\vert 0 \rangle \rightarrow \vert -1 \rangle$ ESR transitions; the matrix element coupling these transitions is thus sensitive to the polarization of dipole fields emanating from the magnet. Particularly, for circularly polarized dipole fields (i.e. near $e_{\rm QI} \sim 1$) only one handedness can cause ESR transitions (c.f. Fig.~\ref{schematic_and_chiral}d ). 

The handedness of nanomagnet's dipole field is, in turn, governed by the equilibrium orientation of magnetization, reversing sense for opposite orientations (see Fig.~\ref{schematic_and_chiral}c). To corroborate the chiral coupling origin of the observed ellipticity dependence, we thus also calculate $\Gamma_\pm$ for the case where the equilibrium orientation of the nanomagnet is flipped. Indeed, as seen in Fig.~\ref{schematic_and_chiral}c, now around $e_{\rm QI} \sim 1$: $\Gamma_-$ becomes negligible while $\Gamma_+$ remains large throughout. Interestingly, this suggests that one of the ESR transitions can be decoupled from the nanomagnet's noise by designing the mode ellipticity. Moreover, the QI spin-nanomagnet coupling for a particular ESR transition can be turned on and off by switching the orientation of the nanomagnet, adding the unique knob of dynamical tunability to nanomagnet-QI hybrids.

\textit{Coherent driving via small-angle dynamics}| Next, we turn to study coherent electrical driving of the QI spin in nanomagnet-QI hybrids by leveraging the electrical activation of a nanomagnet's dynamical modes. In particular, we focus here on electric-field driving of nanomagnet (as opposed to current-induced driving), since such a scheme requires orders of magnitude less power \cite{nozaki2012electric}, and would thus introduce minimal heating-induced decoherence. To do so, we propose to leverage the nanomagnet/oxide heterostructures, where the coupling of electric field $E$ (or equivalently voltage $V$ across the heterostructure) to the magnetization gives rise to following additional term in the magnet's free energy: $\mathcal{F_{\rm m-E}}= M_s \xi V m_z^2/2 $ (see Fig.~\ref{ParametricLAD}a). Such a term can be understood as voltage control of magnetic anisotropy (VCMA) \cite{Zhu2012VCMA,nozaki2012electric}, which arises due to electric field-induced modification of interfacial orbital occupation in combination with spin-orbit interaction. Here, $\xi = 2 \beta/(M_s t_{ox} d_F)$ with $\beta$ being the VCMA coefficient and $d_F$ and $t_{ox}$ being the magnet's and oxide's thickness, respectively. 

In analogy with the spin-wave driving of QI spin in the extended film geometry \cite{andrich2017long}, we begin by focusing on the scheme where the QI spin is driven by the electrically pumped \textit{small-angle} magnetization dynamics in nanomagnets. Namely, within this scheme, the voltage-induced torques (arising from VCMA) coherently excites the mode with frequency close to the FMR and precession angles much smaller than 1 radian. This produces coherent dipolar field at the QI, which would induce Rabi-oscillation of the QI-spin. For efficient driving, such a scheme requires to satisfy the resonant condition $\omega_{\rm FMR} \approx \omega_{q}$, where $\omega_{q}$ denotes the splitting of the QI's ESR transition that one wants to drive. 

For simplicity, we fix the geometry to that of an easy-plane nanomagnet ($H_k=0$), where the equilibrium magnetization and QI-spin are oriented along an external field applied in the easy plane ($z$ axis; see Fig~\ref{ParametricLAD}a). In this case, as has been demonstrated experimentally \cite{chen2017parametric}, the FMR mode can be excited parametrically by application of an ac voltage at twice the FMR frequency ($V(t) = V_0 \sin(2\omega_\mrm{FMR} t)$ with amplitude $V_0$ and  frequency $2\omega_\mrm{FMR} $). For parametric driving, the voltage amplitude is required to be above a threshold $V_\mrm{th} = \alpha \omega_\mrm{FMR}/\eta$ to host a finite magnetization deviation amplitude where $\eta = \gamma \xi \omega_1/(4\omega_\mrm{FMR})$ \cite{verba2014para}. Above this threshold, the inherent nonlinearity of the LLG equation prevents the amplitude of magnetization deviation from diverging \cite{chen2017parametric}. 

The nonlinearity can be evaluated by solving the LLG beyond the linear order and is found to be $\Psi = \gamma[H_k (\omega_\mrm{FMR}/\omega_1) + (H_k + H_\perp) (\omega_1/\omega_\mrm{FMR})]/2$ \cite{supp_ref}. Correspondingly, the amplitude of magnetization deviation in the two transverse directions to the equilibrium is given by $ \vert \delta m'_x \vert = \vert c \vert \vert \delta m'_{0,x} \vert $ and $ \vert \delta m'_y \vert = \vert c \vert \vert \delta m'_{0,y} \vert $, where $\vert c \vert^2 = \eta \sqrt{V_0^2 - V_\mrm{th}^2}/\Psi$ \cite{chen2017parametric}. For the given geometrical factors, $\bar{\mathcal{B}}_x^y = \bar{\mathcal{B}}_y^x \approx 0$ and thus, the corresponding transverse field component amplitude at the QI are $\vert H^{QI}_{0,x}\vert = \vert \bar{\mathcal{B}}_x^x \delta m'_x \vert$ and $\vert H^{QI}_{0,y} \vert = \vert \bar{\mathcal{B}}_y^y \delta m'_y \vert$. This coherent field from the nanomagnet causes Rabi oscillations with frequency $\Omega_R = \gamma[\vert H^{QI}_{0,x}\vert + \vert H^{QI}_{0,y}\vert]/(\sqrt{2})$. For typical nanomagnets-QI parameters and VCMA parameters (see caption of Fig.~\ref{ParametricLAD}), this Rabi frequency corresponds to $\Omega_R/(2\pi) \approx 40$ MHz.  

While selecting $\omega_{\rm FMR} \approx \omega_q$ maximizes Rabi frequency, as discussed in the QI-relaxation section, the same condition also efficiently couples incoherent fields  emanating from the magnet to QI, causing decoherence (see Fig.~\ref{ParametricLAD}b). Using $\Gamma_-$ from Eq.~\ref{RateEq} as the scale for decoherence rate, for the geometrical and material parameters considered [c.f. Fig.~\ref{ParametricLAD}], the estimated order of room temperature quality factor $Q = \Omega_R/\Gamma_- \approx 10$. 

To test the feasibility of the proposed scheme, we numerically solve the s-LLG equation (including VCMA-induced and thermal torques) coupled with the master equation for the density matrix associated with the QI $\rho$: $\dot{\rho} = -i [\mathcal{H}_{QI}^- +\mathcal{H}_{m-QI}^-,\rho]$ ( see supplementary \cite{supp_ref} for the details). The obtained numerical results are shown in Fig.~\ref{ParametricLAD}b averaged over 500 realizations of the stochastic thermal field. While coherent oscillations are observed, the oscillation frequency is of the same order as that of decoherence rate. We note that the corresponding numerically obtained $Q$ is smaller than the analytical estimate. This  may result from paramateric drive-induced enhancement of thermal fluctuations \cite{Li2020spinphonon}. Nevertheless, both analytical result and numerical simulations suggest that, while using the small-angle dynamics it is possible to drive coherent Rabi oscialltion of QI spin electrically at room temperature, the corresponding small quality factors may not be sufficient for some applications. Q could be improved by enhancing the VCMA coefficient, which would increase the strength of coherent fields for same applied voltage, and/or by using off-resonant driving schemes to reduce the impact of thermal decoherence. We next show that the inherent non-linear precessional dynamics of nanomagnet presents an alternate resource to achieve this goal. 

\textit{Coherent driving via large-angle dynamics}| Due to the absence of Suhl instabilities \cite{andersonsuhl1955,suhl1957theory}, nanomagnets can host unique large angle dynamics non-linear precessional modes. These modes are naturally separated in frequency from thermally driven FMR mode and can be selectively excited by the electrical drives \cite{shiota2012LAD} (but not just by thermal noise). At the same time, these large angle dynamics mode cause larger amplitude dipolar fields at the QI, which increases the Rabi frequency. Thus, they offer the exciting opportunity to reduce coupling to thermal noise while efficiently coupling coherent electrical signals. We demonstrate next high quality factor driving of QI spins taking advantage of electrically activated nonlinear precessional mode of the nanomagnet (referred to as LAD-scheme).

To illustrate LAD-scheme, we focus on the case of an out-of-plane nanomagnet ($H_\perp<0$), which can be attained, for example, by modulating the interfacial anisotropy in nanomagnet/oxide heterostructure via controlling the magnet's thickness \cite{ikeda2010perpendicular}. In this case, an in-plane external field applied along the z-axis (the axis along which the QI also points) tilts the magnet's equilibrium in the xz-plane at an angle $\cos\theta_0 = H_\mrm{ext}/\vert H_\perp \vert$ for $H_\mrm{ext}<\vert H_\perp \vert$ (see Fig.~\ref{ParametricLAD}d). If a voltage pulse is applied to turn off the effective perpendicular anisotropy i.e. $H_\perp(V_0)=0$ (when $V_0 = H_\perp/\xi$), the magnetization experiences only the external field and precesses about it at a frequency $\omega_{\rm LAD} \sim \gamma H_{\rm ext}$. Such voltage-induced precessional dynamics have been demonstrated experimentally, attracting significant interest for constructing all electrically controlled ultra-low energy toggle magnetic memory \cite{shiota2012LAD}. For our purpose, such precessions produces coherent dipolar fields at the QI, which when resonant with the $\vert 0 \rangle \leftrightarrow \vert -1 \rangle$ ESR transition of QI (i.e. $\omega_{\rm LAD}=\omega^{QI}_-$) induces Rabi-oscillations. 

Crucially, in the absence of voltage driving, the magnet cannot perform the large angle dynamics; instead it is fluctuating near its equilibrium due to thermal noise, and producing incoherent fields at the QI with spectral density peaked at $\omega_{\rm FMR}\sim \sqrt{\omega_1 \omega_2}$. Consequently, by being in the regime of $\omega_{\rm LAD}=\omega_q \neq \omega_{\rm FMR}$, the nanomagnet's voltage-induced large angle coherent dynamics can be coupled efficiently to the QI spin, while the coupling to thermal fluctuations-induced incoherent dynamics is reduced. 

Motivated by this, we present the pulse sequence for the LAD-scheme in Fig.~\ref{ParametricLAD}d. We use the voltage pulse sequence $V(t) = V_0 \sum_{n=0}^{N}[\Theta(t-n(\tau_p+\tau_w))-\Theta(t-n(\tau_p+\tau_w)-\tau_p)]$ where $\Theta$ is the Heaviside step function, $N$ is the maximum number of pulses applied, $\tau_p$ and $\tau_w$ are the time duration's for which the voltage is turned on and off, respectively. We let the voltage pulse on for a time equal to the time for one full magnetization precession i.e. $\tau_p = 2\pi/(\gamma H_\mrm{ext})$; keeping the pulse on for longer time would drive the magnet towards the new equilibrium (along $z$), which would eventually reduce the amplitude of dynamic coherent fields to zero. Furthermore, we choose a wait time $\tau_w = \gamma H^{QI}_\mrm{eff} = \gamma \sqrt{(\Delta/\gamma - H_\mrm{ext} - H^{QI}_\mrm{eq,z})^2 + 2 (H^{QI}_\mrm{eq,x})^2}$ such that the azimuthal angle of the QI spin state is at the same angle at the beginning of each voltage pulse. Note that $H^{QI}_\mrm{eq,i}$ ($i=\{x,y\}$) are the equilibrium field components at the QI from the equilibrium magnetization causing an effective field $H^{QI}_\mrm{eff}$ about which the QI spin state precesses. 

To test LAD-scheme, we implement it numerically by solving the coupled s-LLG equation (with time-dependent $H_\perp$ and stochastic thermal field) and the master equation for the QI (influenced by $\mathcal{H}_{m-QI}^-$), averaging over 500 realizations of the stochastic thermal field. During the time $\tau_p$, the voltage turns off the anisotropy and causes the magnetization to precess about the external field, causing dipolar field at the QI with a handedness that induces Rabi dynamics between the $\vert 0 \rangle \leftrightarrow \vert -1\rangle$ levels (c.f. Fig.~\ref{ParametricLAD}e). As seen from Fig.~\ref{ParametricLAD}e, the quality factor $Q \approx 1400$ at room temperature \cite{ladQcomment}. Remarkably, for similar driving voltage amplitudes, the LAD-driven scheme enhances room-temperature $Q$ by about \textit{two orders} of magnitude in comparison to the case of driving the QI via the FMR mode (c.f Fig.~\ref{ParametricLAD}e). A close up look at the QI spin state dynamics shows step-like change in probability which is due to the fact that during the time $\tau_p$ over which the magnetization undergoes one large precession, the consequent rotating dipolar field at the QI causes Rabi dynamics.

\begin{figure}[hbtp]
\centering
\includegraphics[width=0.49\textwidth]{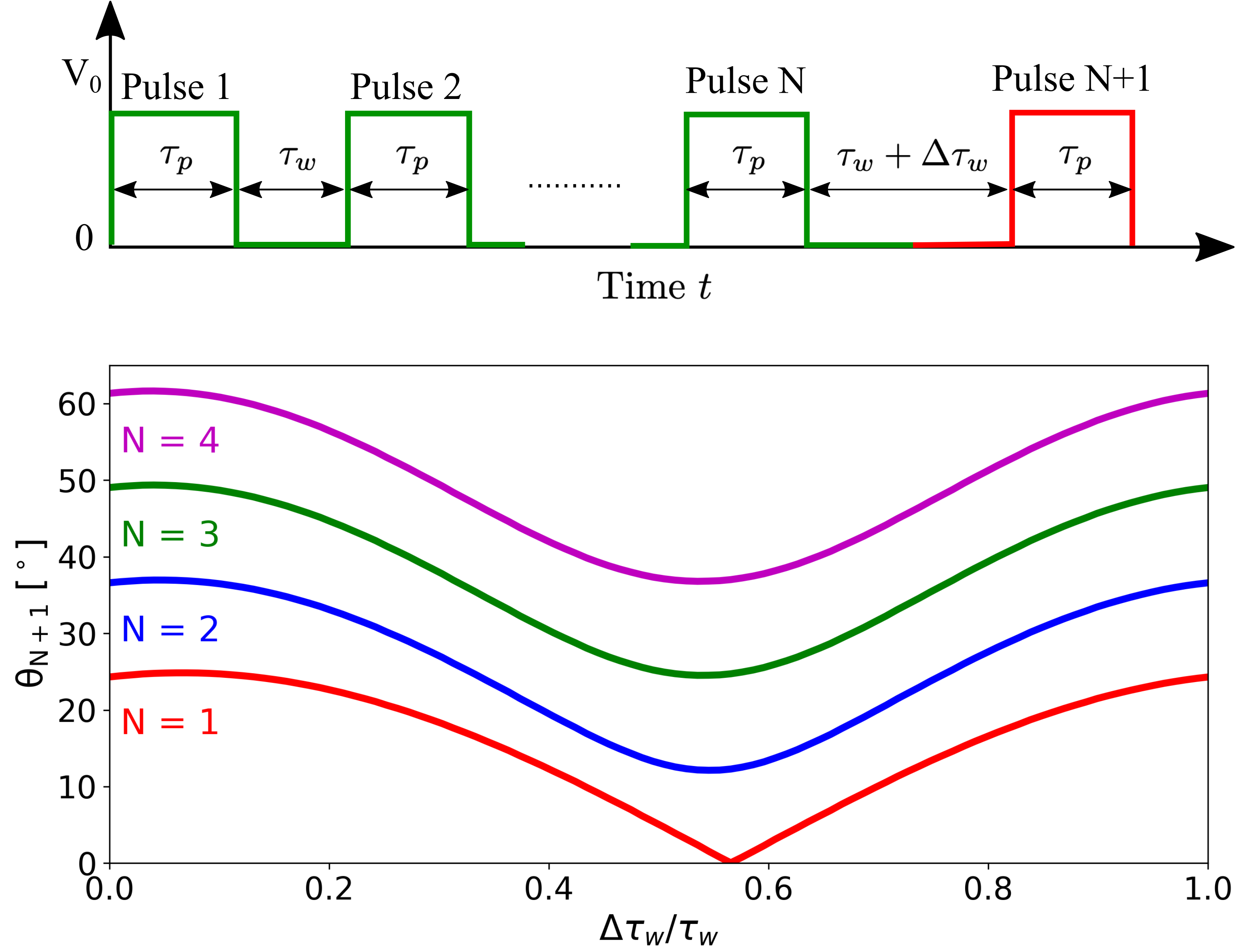}
\caption{\label{Rabi2} Coherent driving of the QI spin to arbitrary polar angle over the Bloch sphere by appropriately timing an additional voltage pulse at a delay time $\tau_w + \Delta\tau_w$ after the $N^\mrm{th}$ pulse where a) shows the time-dependent voltage pulse, and b) shows the consequent final Bloch sphere polar angle.}
\end{figure}

\textit{Arbitrary QI-spin rotation via large-angle dynamics}| For a complete control, it is imperative for the drive to change the state of the QI-qubit to any arbitrary point on the Bloch sphere.  In Fig.~\ref{Rabi2}, we show that for the LAD-scheme it is also possible to reach an arbitrary polar angle. For this purpose, we take advantage of the fact that the amount by which the QI-qubit rotates on the Bloch sphere is controlled by the phase of the Rabi-driving field \cite{krantz2019quantum}, which can be understood as follows. In the absence of a drive field, the QI-qubit performs free Larmor precessions about the quantization axis, which keeps changing the azimuthal angle on the Bloch sphere. Thus, the relative initial angle between the QI-qubit and the drive field, which decides the change in the polar angle on the Bloch sphere, is controlled by the  azimuthal location of QI-qubit when the drive is turned on.

The phase of the coherent dipole field emanating from the magnet can, in turn, be controlled by adjusting the time at which a voltage pulse is applied; as the magnet only performs the large angle precession, and consequently produces a coherent dipole field, for the time when the voltage pulse is on. To this end, we add a time $\Delta \tau_w$ between voltage pulse $N$ and $N+1$ in the numerical simulations and plot in Fig.~\ref{Rabi2} the final polar angle, $\theta_{N+1}$, reached after the N+1 pulse. When $\Delta \tau_w=0$, the voltage sequence becomes same as that of Fig~\ref{ParametricLAD}d, giving the same discrete values of $\theta_{N+1}$. On the other hand, by smoothly varying $\Delta \tau_w$ from zero to $t_w$, we see that any arbitrary $\theta_{N+1}$ can be reached. Combined with free Larmor precessions, this demonstrates that the electrically controlled nanomagnet's LAD dynamics can be leveraged as a coherent drive for arbitrary rotation of the QI's spin-state over the Bloch sphere.

\textit{Conclusion}| In summary, we propose and study theoretically quantum impurity spins coupled with the dynamical modes of nanomagnets, which are controlled via thermal and/or electrical torques. We provide a quantitative understanding of the thermally excited mode-induced QI-spin decoherence in such hybrid system. We find that the decoherence is sensitive to the mode ellipticity and the inherent chiral nature of coupling between QI spin and the magnon mode. This allows for tuning the decoherence via geometry and material design, along with providing the unique ability to turn it on and off via reorienting magnetic orientation. This understanding can guide future applications using QI-spin decoherence as a resource for sensing. 

However, when coherently driving the QI spin via exploiting electrically driven modes, the magnet-induced decoherence restricts QI-spin manipulation to low quality factors at room-temperature. To circumvent this, we propose and demonstrate a means to drive the QI spin leveraging the nonlinear precessional dynamics of the nanomagnet. Within this scheme, the magnet's large angle dynamics couples efficiently to the QI but the coupling to thermally-induced decoherence is significantly reduced. This improves the quality factor by over two orders of magnitude thus providing a pathway for larger quality factor local electrical driving of QI spins at ambient conditions. 

\vspace{0.1 in}
%\section{Acknowledgements}
\textit{Acknowledgements}| AR and PU acknowledge support by the U.S. Department of Energy, Office of Science through the Quantum Science Center (QSC), a National Quantum Information Science Research Center, NSF ECCS-1944635, and NSF ECCS-1810494.

\bibliography{refs}

\begin{thebibliography}{53}
\bibitem [{sup()}]{supp_ref}%
  \BibitemOpen
  \href@noop {} {}\bibinfo {note} {See Supplemetal Material at ------------,
  where we present a detailed derivation of the nanomagnet induced QI
  relaxation, magnetization deviation amplitude via parametric driving, and
  numerical details.}\BibitemShut {Stop}%
\bibitem [{lad()}]{ladQcomment}%
  \BibitemOpen
  \href@noop {} {}\bibinfo {note} {The relaxation time is extracted from
  Fig.~\ref{ParametricLAD}e by fitting a decaying exponential and the Rabi
  dynamics time period is read off from the close-up plot of the TLS
  dynamics.}\BibitemShut {Stop}%
\end{thebibliography}

%%%% Supplementary Material %%%%%%%%%%%
%%%%%%%%%%%%%%%%%%%%%%%%%%%%%%%%%%%%%%%
\clearpage

\setcounter{equation}{0}
\setcounter{figure}{0}
\setcounter{table}{0}
\setcounter{page}{1}
\setcounter{section}{0}
% \makeatletter
\renewcommand{\theequation}{S.\arabic{equation}}
\renewcommand{\thefigure}{S\arabic{figure}}
\renewcommand{\bibnumfmt}[1]{[S#1]}
\renewcommand{\citenumfont}[1]{S.#1}
\renewcommand{\citenumfont}[1]{\textit{#1}}

\onecolumngrid

\begin{center}
\textbf{\large Supplemental Material}
\end{center}

In this supplemental material, we provide a detailed derivation of the theoretical formalism used for evaluating the quantum impurity relaxation rates caused by dipolar field fluctuations emanating from a nanomagnet in thermal equilibrium. We next derive the nanomagnet eigenmodes and the amplitude of magnetization deviations under parametric driving. We end by providing the details of solving for the dynamics of the QI-nanomagnet hybrid.

\section{Quantum Impurity Relaxometry}
In a prototypical spin-1 quantum impurity with a spin triplet ground state labeled by the projection quantum numbers $m_s = \{-1,0,1\}$ along the quantization axis, the zero field splitting $\Delta$ (typically present in the QIs with spin $>$ 1/2) splits the otherwise degenerate $m_s$ levels to a lower energy $m_s=0$ and the higher energy degenerate $m_s=\{-1,1\}$ levels. The Hamiltonian governing the dynamics of the QI in presence of a magnetic field $\vec{H}^{QI}$ is thus given by
\begin{equation}
\mathcal{H}_{QI} = \Delta \, S_z^2 + \gamma \, \vec{S}\cdot \vec{H}^{QI} =  \Delta \, S_z^2 + \gamma \, \left[ S_x H_x^{QI} + S_y H_y^{QI} +S_z H_z^{QI}\right],
\end{equation} 
where $\gamma>0$ is the gyromagnetic ratio and the spin-1 matrices are
\begin{equation}
S_z = \left(\begin{array}{ccc}
1 & 0 & 0   \\ 
0 & 0 & 0   \\ 
0 & 0 & -1   
\end{array} \right) \quad S_x = \dfrac{1}{\sqrt{2}} \left(\begin{array}{ccc}
0 & 1 & 0   \\ 
1 & 0 & 1   \\ 
0 & 1 & 0   
\end{array} \right) \quad S_y = \dfrac{1}{\sqrt{2}} \left(\begin{array}{ccc}
0 & -i & 0   \\ 
i & 0 & -i   \\ 
0 & i & 0   
\end{array} \right).
\end{equation}
Lifting the degeneracy between the $m_s=\pm 1$ states in the above Hamiltonian by the application of a dc magnetic field along the QI quantization axis allows us to treat the three level system as two effective two-level systems (TLSs) labeled by superscripts `+' (formed by $m_s$ = 0 and $m_s$=+1) and `-' (formed by $m_s$ = 0 and $m_s$=-1), respectively. As shown in [Ref. S1], the Hamiltonians corresponding to the effective TLSs can be written in a concise form as
\begin{equation}
\label{supp_NVHamiltonian}
\begin{split}
\mathcal{H}^{\pm}_{QI,TLS} &=\dfrac{\omega_\pm^{QI}}{2} [\mathcal{I} \pm \sigma_z] + \dfrac{\gamma}{2\sqrt{2}} [H_+^{QI} \sigma_- + H_-^{QI} \sigma_+], \\
\end{split}
\end{equation}
where $\omega_\pm^{QI}$ is the energy separation between the states forming the effective TLSs, $H_\pm^{QI} = H_x^{QI} \pm i H_y^{QI}$, $\sigma_\pm = \sigma_x \pm i \sigma_y$, and $\sigma_i$ ($i-\{x,y,z\}$) is the Pauli matrix.\\

The fluctuating magnetic field components transverse to the quantization axis causes relaxation of an initialized QI. Thus, we define the effective Hamiltonian as a sum of the equilibrium and perturbation terms,
\begin{equation}
\mathcal{H}_\mrm{TLS}^{\pm} = \mathcal{H}_{QI,0}^{\pm} + V(t),
\end{equation}
where $\mathcal{H}_{QI,0}^{\pm} = \dfrac{\omega_\pm^{QI}}{2} [\mathcal{I} \pm \sigma_z]$ and $V(t) = \dfrac{\gamma}{2\sqrt{2}} [H_+^{QI} \sigma_- + H_-^{QI} \sigma_+]$. The rate of transition from the $\vert 0 \rangle$ to $\vert \pm 1 \rangle$ state is given by [Ref. S1,S2,S3]
\begin{equation}
\label{supp_rate_exp}
\Gamma_\pm (\omega_\pm^{QI}) = \dfrac{\gamma^2}{2} \int dt \, e^{i \omega_\pm^{QI} t} \langle H_\mp^{QI}(t) H_\pm^{QI} (0) \rangle,
\end{equation}
written as an inline equation in the `Relaxation Model' section of the main manuscript.\\

Since we are interested in evaluating the fluctuating magnetic field components arising from the magnetic moment fluctuations in the nanomagnet, we relate this relaxation rate to the incoherent dynamics of the nanomagnet in thermal equilibrium.

\section{Incoherent nanomagnet dynamics}
\textit{Free Energy}| The free energy of the nanomagnet with in-plane easy axis (along z-axis), (shape+interfacial) perpendicular anisotropy (along x-axis), and external field in the xz-plane (at an angle $\theta_H$ with respect to the z-axis) is 
\begin{equation}
\mathcal{F} = -M_s \vec{m}\cdot \vec{H}_\mrm{ext} - K m_z^2 +\dfrac{M_s}{2} H_\perp m_x^2.
\end{equation}

\textit{Equilibrium Magnetization}| Given the free energy, the equilibrium magnetization is oriented at an angle that minimizes the free energy. For generality, we denote this angle as $\theta$ which is measured relative to the z-axis in the x-z plane.

\textit{Eigenmode}| To solve for the FMR mode frequency and eigenvector, we linearize the Landau-Lifshitz equation [Ref S4] about the magnet equilibrium. Thus, it makes sense to use Landau-Lifshitz equation in the magnet frame (where the z-axis aligns along the magnet equilibrium - achieved by rotating by an angle $\theta_0$ about the y-axis)
\begin{equation}
\dfrac{d \vec{m}'}{dt} = - \gamma \, \vec{m}' \times \vec{H}'_\mrm{eff}.
\end{equation}
Using
\begin{equation}
\vec{m} = R_y(\theta_0) \vec{m}' = \left(\begin{array}{c}
m'_x \cos\theta_0 + m'_z \sin\theta_0 \\ 
m'_y \\ 
-m'_x \sin\theta_0 + m'_z \cos\theta_0 
\end{array} \right),
\end{equation}
the effective field in the magnet frame is
\begin{equation}
\vec{H}'_\mrm{eff} =  \left(\begin{array}{c}
H_0 \sin(\theta_H - \theta_0) + m'_x (H_k \sin^2\theta_0 - H_\perp \cos^2\theta_0) - m'_z (H_k+H_\perp) \sin\theta_0 \cos \theta_0\\ 
0 \\ 
H_0 \cos(\theta_H - \theta_0) - m'_x (H_k+H_\perp) \sin\theta_0 \cos \theta_0+ m'_z (H_k \cos^2\theta_0 - H_\perp \sin^2\theta_0) 
\end{array} \right). 
\end{equation}
Linearizing the Landau-Lifshitz equation
\begin{equation}
\dfrac{d \delta \vec{m}'}{dt} = - \gamma \, \vec{m}'_\mrm{eq} \times \delta\vec{H}'_\mrm{eff}  - \gamma \, \delta\vec{m}' \times \vec{H}'_\mrm{eff,eq},
\end{equation}
where
\begin{equation}
\vec{H}'_\mrm{eff,eq} =  \left(\begin{array}{c}
H_0 \sin(\theta_H - \theta_0) -  (H_k+H_\perp) \sin\theta_0 \cos \theta_0\\ 
0 \\ 
H_0 \cos(\theta_H - \theta_0) + H_k \cos^2\theta_0 - H_\perp \sin^2\theta_0
\end{array} \right) \equiv   \left(\begin{array}{c}
H'_{x,\mrm{eq}}\\ 
0 \\ 
H'_{z,\mrm{eq}}
\end{array} \right)  \qquad \delta\vec{H}'_\mrm{eff} =  \left(\begin{array}{c}
\delta m'_x (H_k \sin^2\theta_0 - H_\perp \cos^2\theta_0) \\ 
0 \\ 
- \delta m'_x (H_k+H_\perp) \sin\theta_0 \cos \theta_0
\end{array} \right) 
\end{equation}
and
\begin{equation}
\vec{m}'_\mrm{eq} =  \left(\begin{array}{c}
0\\ 
0 \\ 
1
\end{array} \right)  \qquad \delta\vec{m}' =  \left(\begin{array}{c}
\delta m'_x  \\ 
\delta m'_y  \\ 
0
\end{array} \right) .
\end{equation}
Therefore,
\begin{equation}
\label{supp_LLeq}
\dfrac{d}{dt} \left(\begin{array}{c}
\delta m'_x\\ 
\delta m'_y 
\end{array} \right) = \left( \begin{array}{cc}
0  & -\gamma H'_{z,\mrm{eq}} \\ 
\gamma \left[ H'_{z,\mrm{eq}}+ H_\perp \cos^2\theta_0 - H_k \sin^2\theta_0 \right] &  0 
\end{array} \right) \left(\begin{array}{c}
\delta m'_x\\ 
\delta m'_y 
\end{array} \right).
\end{equation}
Non-trivial solution for the magnetization deviations provides the FMR resonance frequency 
\begin{equation}
\omega_\mrm{FMR} = \gamma \sqrt{\left[ H_0 \cos(\theta_H - \theta_0) + H_k \cos^2\theta_0 - H_\perp \sin^2\theta_0\right]\left[ H_0 \cos(\theta_H - \theta_0) + (H_k + H_\perp) \cos2\theta_0\right]}
\end{equation},
and the associated eigenvector
\begin{equation}
\delta\vec{m}^\prime_0 = \dfrac{1}{\sqrt{\omega_\mrm{FMR}^2 + \gamma^2 H_{z,\mrm{eq}}^{'2}}} \left( \begin{array}{c}
\gamma H'_{z,\mrm{eq}}  \\ 
-i \omega_\mrm{FMR} 
\end{array} \right).
\end{equation}

\textit{Susceptibility}| To get the magnetic susceptibility response to an small applied oscillatory field $\vec{h}'_\sim$, we can again look at the linearized Landau-Lifshitz equation but now include the Gilbert damping (i.e. LLG equation) [Ref. S4, S5] 
\begin{equation}
\dfrac{d \delta \vec{m}'}{dt} = - \gamma \, \vec{m}'_\mrm{eq} \times \delta\vec{H}'_\mrm{eff}  - \gamma \, \delta\vec{m}' \times \vec{H}'_\mrm{eff,eq} - \gamma \, \vec{m}'_\mrm{eq} \times \delta\vec{h}'_\sim + \alpha \, \vec{m}'_\mrm{eq} \times \dfrac{d \delta \vec{m}'}{dt}.
\end{equation}
Defining frequency components 
\begin{equation}
\begin{split}
\omega_1 &= \gamma \left[ H_0 \cos(\theta_H - \theta_0) + H_k \cos^2\theta_0 - H_\perp \sin^2\theta_0 \right] \\
\omega_2 &= \gamma \left[ H_0 \cos(\theta_H - \theta_0) + (H_k + H_\perp) \cos2\theta_0\right]
\end{split}
\end{equation}
and assuming harmonic time dependence for $\vec{h}'_\sim \sim e^{-i\omega t}$ and $\delta\vec{m}' \sim e^{-i\omega t}$,
\begin{equation}
\delta m'_x = S'_{xx} h'_x + S'_{xy} h'_y \qquad \delta m'_y = S'_{yx} h'_x + S'_{yy} h'_y,
\end{equation}
where the susceptibilities are
\begin{equation}
\begin{split}
S'_{xx} &= \dfrac{-\gamma [\omega_1 - i \alpha \omega]}{\omega^2 -[\omega_1 - i \alpha \omega][\omega_2 - i \alpha \omega]} \\
S'_{xy} &= \dfrac{i \gamma \omega }{\omega^2 -[\omega_1 - i \alpha \omega][\omega_2 - i \alpha \omega]} \\
S'_{yx} &= \dfrac{-i \gamma \omega }{\omega^2 -[\omega_1 - i \alpha \omega][\omega_2 - i \alpha \omega]} \\
S'_{yy} &= \dfrac{-\gamma [\omega_2 - i \alpha \omega]}{\omega^2 -[\omega_1 - i \alpha \omega][\omega_2 - i \alpha \omega]}. \\
\end{split}
\end{equation}

\textit{Magnetization Correlations}| The fluctuating fields $\vec{h}'$ whose correlators are known,
\begin{equation}
\langle h'_i (t) h'_j(t^\prime) \rangle=2D_\mathrm{th} \delta_{ij}\delta(t-t^\prime) \quad \Rightarrow \quad  \langle h'_i (\omega) h'_j(\omega^\prime) \rangle=2\pi \, 2D_\mathrm{th} \delta_{ij}\delta(\omega+\omega^\prime),
\end{equation}
where $D_\mathrm{th} = \alpha k_B T/ \gamma M_s v $ ($v$ is the nanomagnet volume) [Ref. S6], are useful in finding correlation functions of the form
\begin{equation}
\langle \delta m'_{i,inc} (t) \delta m'_{j,inc}(0) \rangle = \int \dfrac{d\omega}{2\pi}\,\int \dfrac{d\omega^\prime}{2\pi}\, e^{-i\omega t}\langle \delta m'_{i,inc} (\omega) \delta m'_{j,inc}(\omega^\prime) \rangle. 
\end{equation}
Using the ac-field fluctuator correlators, the magnetization correlators are related to the susceptibilities
\begin{equation}
\begin{split}
\langle  \delta m'_{x,inc} (\omega) \delta m'_{x,inc}(\omega^\prime) \rangle &=2\pi \, 2D \delta(\omega+\omega^\prime)  \left[ S'_{xx}(\omega) S'_{xx}(-\omega) + S'_{xy}(\omega) S'_{xy}(-\omega) \right]   \\
\langle  \delta m'_{y,inc} (\omega) \delta m'_{y,inc}(\omega^\prime) \rangle &= 2\pi \, 2D \delta(\omega+\omega^\prime)  \left[ S'_{yx}(\omega) S'_{yx}(-\omega) + S'_{yy}(\omega) S'_{yy}(-\omega) \right]   \\
\langle  \delta m'_{x,inc} (\omega) \delta m'_{y,inc}(\omega^\prime) \rangle &= 2\pi \, 2D \delta(\omega+\omega^\prime)  \left[ S'_{xx}(\omega) S'_{yx}(-\omega) + S'_{xy}(\omega) S'_{yy}(-\omega)\right]   \\
\langle  \delta m'_{y,inc}(\omega) \delta m'_{x,inc} (\omega^\prime) \rangle &= 2\pi \, 2D \delta(\omega+\omega^\prime)  \left[ S'_{yx}(\omega) S'_{xx}(-\omega) + S'_{yy}(\omega) S'_{xy}(-\omega) \right].   \\
\end{split}
\end{equation}
Given that the system has time-translation invariance,
\begin{equation}
\begin{split}
C'_{ij}(t) &= \int \dfrac{d\omega}{2\pi} \, e^{-i\omega t} \, 2D \left[ S'_{ix}(\omega) S'_{jx}(-\omega) + S'_{iy}(\omega) S'_{jy}(-\omega) \right].  \\
\end{split}
\end{equation}
Thus
\begin{equation}
\begin{split}
C'_{ij}(\omega) =  2D \left[ S'_{ix}(\omega) S'_{jx}(-\omega) + S'_{iy}(\omega) S'_{jy}(-\omega) \right].  \\
\end{split}
\end{equation}
The various correlation are therefore given by
\begin{equation}
\begin{split}
C'_{xx}(\omega) &=  2D \gamma^2 \dfrac{\omega_1^2 + \omega^2 (1+\alpha^2)}{[\omega^2 (1+\alpha^2) - \omega_1 \omega_2]^2 + \alpha^2 \omega^2 (\omega_1+\omega_2)^2}  \\
C'_{yy}(\omega) &=  2D \gamma^2 \dfrac{\omega_2^2 + \omega^2 (1+\alpha^2)}{[\omega^2 (1+\alpha^2) - \omega_1 \omega_2]^2 + \alpha^2 \omega^2 (\omega_1+\omega_2)^2}  \\
C'_{xy}(\omega) &=  2D \gamma^2 \dfrac{-i \omega (\omega_1+ \omega_2)}{[\omega^2 (1+\alpha^2) - \omega_1 \omega_2]^2 + \alpha^2 \omega^2 (\omega_1+\omega_2)^2}  \\
C'_{yx}(\omega) &=  2D \gamma^2 \dfrac{i \omega (\omega_1 + \omega_2)}{[\omega^2 (1+\alpha^2) - \omega_1 \omega_2]^2 + \alpha^2 \omega^2 (\omega_1+\omega_2)^2}.  \\
\end{split}
\end{equation}

We can re-express the magnetization deviation correlators in terms of the magnetization precession ellipticity defined as
\begin{equation}
e = \dfrac{\vert \delta m'_{0,x} \vert}{\vert \delta m'_{0,y} \vert} = \dfrac{\omega_1}{\sqrt{\omega_1 \omega_2}} = \sqrt{\dfrac{\omega_1}{\omega_2}}.
\end{equation}
Note that $e=1$ implies circular magnetization precession, $e=0$ implies linear magentization oscillation along the y-axis, and $e=\infty$ implies linear magentization oscillation along the x-axis. Since $\omega_1 = \omega_\mrm{FMR} e$ and $\omega_2 = \omega_\mrm{FMR}/e$,
\begin{equation}
\begin{split}
C'_{xx}(\omega) &=  2D \gamma^2 \dfrac{\omega_\mrm{FMR}^2 e^2 + \omega^2 (1+\alpha^2)}{[\omega^2 (1+\alpha^2) -\omega_\mrm{FMR}^2 ]^2 + \alpha^2 \omega^2 \omega_\mrm{FMR}^2 (e+1/e)^2}  \\
C'_{yy}(\omega) &=  2D \gamma^2 \dfrac{(\omega_\mrm{FMR}^2/e^2) + \omega^2 (1+\alpha^2)}{[\omega^2 (1+\alpha^2) -\omega_\mrm{FMR}^2]^2 + \alpha^2 \omega^2 \omega_\mrm{FMR}^2 (e+1/e)^2 }  \\
C'_{xy}(\omega) &=  2D \gamma^2 \dfrac{-i \omega \omega_\mrm{FMR} (e+1/e)}{[\omega^2 (1+\alpha^2) - \omega_\mrm{FMR}^2]^2 + \alpha^2 \omega^2 \omega_\mrm{FMR}^2 (e+1/e)^2}  \\
C'_{yx}(\omega) &=  2D \gamma^2 \dfrac{i \omega \omega_\mrm{FMR} (e+1/e)}{[\omega^2 (1+\alpha^2) - \omega_\mrm{FMR}^2]^2 + \alpha^2 \omega^2 \omega_\mrm{FMR}^2 (e+1/e)^2}.  \\
\end{split}
\end{equation}
When on-resonance $\omega = \omega_\mrm{FMR}$,
\begin{equation}
\begin{split}
C'_{xx}(\omega= \omega_\mrm{FMR}) &=  \dfrac{2D \gamma^2}{\alpha^2 \omega_\mrm{FMR}^2} \dfrac{ (1+\alpha^2 + e^2)}{\alpha^2 +  (e+1/e)^2}  \\
C'_{yy}(\omega= \omega_\mrm{FMR}) &=   \dfrac{2D \gamma^2}{\alpha^2 \omega_\mrm{FMR}^2} \dfrac{ (1+\alpha^2 + 1/e^2)}{\alpha^2 + (e+1/e)^2 }  \\
C'_{xy}(\omega= \omega_\mrm{FMR}) &=   \dfrac{2D \gamma^2}{\alpha^2 \omega_\mrm{FMR}^2}  \dfrac{-i  (e+1/e)}{\alpha^2 +  (e+1/e)^2}  \\
C'_{yx}(\omega= \omega_\mrm{FMR}) &=  \dfrac{2D \gamma^2}{\alpha^2 \omega_\mrm{FMR}^2} \dfrac{i  (e+1/e)}{\alpha^2 + (e+1/e)^2}.  \\
\end{split}
\end{equation}

\section{QI relaxation caused by incoherent nanomagnet dynamics}
\textit{Equilibrium field from nanomagnet at the QI}| Given the equilibrium magnetization, there is an equilibrium field at the QI, given by
\begin{equation}
\vec{H}_\mrm{QI,eq} = \left(\begin{array}{ccc}
B_x^x & B_x^y & B_x^z   \\ 
B_y^x & B_y^y & B_y^z   \\ 
B_z^x & B_z^y & B_z^z   
\end{array} \right) \left(\begin{array}{c}
m_{x,\mrm{eq}}\\ 
m_{y,\mrm{eq}} \\ 
m_{z,\mrm{eq}}
\end{array} \right) = \left(\begin{array}{c}
B_x^x \sin\theta_0 + B_x^z \cos\theta_0\\ 
B_y^x \sin\theta_0 + B_y^z \cos\theta_0 \\ 
B_z^x \sin\theta_0 + B_z^z \cos\theta_0
\end{array} \right) \equiv \left(\begin{array}{c}
H_{\mrm{QI,eq},x}\\ 
H_{\mrm{QI,eq},y}  \\ 
H_{\mrm{QI,eq},z}
\end{array} \right),
\end{equation}
where $B_i^j$ accounts for the geometric factors that relate the $j$-th unit magnetization component to the $i$-th field component at QI [Ref. S7]. In the frame of the QI [denoted by the superscript `QI']
\begin{equation}
\vec{H}^{QI}_\mrm{eq} = R_y^T (\theta_\mrm{QI}) \vec{H}_\mrm{QI,eq}   = \left(\begin{array}{c}
H_{\mrm{QI,eq},x} \cos\theta_\mrm{QI} - H_{\mrm{QI,eq},z} \sin\theta_\mrm{QI}\\ 
H_{\mrm{QI,eq},y}  \\ 
H_{\mrm{QI,eq},x} \sin\theta_\mrm{QI} + H_{\mrm{QI,eq},z} \cos\theta_\mrm{QI}
\end{array} \right) \equiv \left(\begin{array}{c}
H^{QI}_{\mrm{eq},x}\\ 
H^{QI}_{\mrm{eq},y}  \\ 
H^{QI}_{\mrm{eq},z}
\end{array} \right),
\end{equation}
where the field components in the QI frame are
\begin{equation}
\begin{split}
H^{QI}_{\mrm{eq},x} & =\cos\theta_\mrm{QI} \left[B_x^x  \sin\theta_0 + B_x^z  \cos \theta_0\right] - \sin\theta_\mrm{QI} \left[ B_z^x \sin\theta_0 + B_z^z \cos\theta_0 \right]\\ 
H^{QI}_{\mrm{eq},y} & = B_y^x \sin\theta_0 + B_y^z \cos\theta_0 \\
H^{QI}_{\mrm{eq},z} & = \sin\theta_\mrm{QI} \left[B_x^x  \sin\theta_0 + B_x^z  \cos \theta_0\right] + \cos\theta_\mrm{QI} \left[ B_z^x \sin\theta_0 + B_z^z \cos\theta_0 \right].
\end{split}
\end{equation}

\textit{Incoherent field from nanomagnet at the QI}| Similar to field from equilibrium magnetization, the dynamic deviations in magnetization lead to dynamic field components at the QI,
\begin{equation}
\vec{H}_\mrm{QI} = \left(\begin{array}{ccc}
B_x^x & B_x^y & B_x^z   \\ 
B_y^x & B_y^y & B_y^z   \\ 
B_z^x & B_z^y & B_z^z   
\end{array} \right) \left(\begin{array}{c}
\delta m_{x}\\ 
\delta m_{y} \\ 
\delta m_{z}
\end{array} \right) = \left(\begin{array}{ccc}
B_x^x & B_x^y & B_x^z   \\ 
B_y^x & B_y^y & B_y^z   \\ 
B_z^x & B_z^y & B_z^z   
\end{array} \right) \left( \begin{array}{ccc}
\cos\theta_0 & 0 & \sin\theta_0 \\ 
0 & 1 & 0 \\ 
-\sin\theta_0 & 0 & \cos\theta_0
\end{array} \right) \delta \vec{m}'.
\end{equation}

In the QI frame (where the z-axis is aligned along the QI quantization axis), 
\begin{equation}
\begin{split}
\vec{H}^{QI} = R_y^T (\theta_\mrm{QI}) \vec{H}_\mrm{QI} &=  \left( \begin{array}{ccc}
\cos\theta_\mrm{QI} & 0 & -\sin\theta_\mrm{QI} \\ 
0 & 1 & 0 \\ 
\sin\theta_\mrm{QI} & 0 & \cos\theta_\mrm{QI}
\end{array} \right) \left(\begin{array}{ccc}
B_x^x & B_x^y & B_x^z   \\ 
B_y^x & B_y^y & B_y^z   \\ 
B_z^x & B_z^y & B_z^z   
\end{array} \right) \left( \begin{array}{ccc}
\cos\theta_0 & 0 & \sin\theta_0 \\ 
0 & 1 & 0 \\ 
-\sin\theta_0 & 0 & \cos\theta_0
\end{array} \right) \delta \vec{m}' \\
&\equiv \left(\begin{array}{ccc}
\bar{B}_x^x & \bar{B}_x^y & \bar{B}_x^z   \\ 
\bar{B}_y^x & \bar{B}_y^y & \bar{B}_y^z   \\ 
\bar{B}_z^x & \bar{B}_z^y & \bar{B}_z^z   
\end{array} \right)  \delta \vec{m}'.
\end{split}
\end{equation}

\textit{Population relaxation rate for the QI}| The Hamiltonians corresponding to the effective TLSs can be written in a concise form as
\begin{equation}
\label{supp_NVHamiltonian2}
\begin{split}
\mathcal{H}^{\xi_\pm}_{TLS} &=\dfrac{\omega^{QI}_{\pm}}{2} [\mathcal{I} \pm \sigma_z] + \dfrac{\gamma}{2\sqrt{2}} [H^{QI}_{+} \sigma_- + H^{QI}_{-} \sigma_+]. \\
\end{split}
\end{equation}
For typical experiments, the external field in a hybrid QI-magnet system is applied along the QI quantization axis. Thus, $\theta_\mrm{QI} = \theta_H$. As a consequence, the ESR frequencies for the effective TLSs is determined by the external field and the component of field produced by the equilibrium magnetization along the QI quantization axis. Hence,
\begin{equation}
\omega^{QI}_{\pm} = \Delta \pm \gamma [H_0 + H^{QI}_{\mrm{eq},z}],
\end{equation}
where $\Delta$ is the zero field splitting. The relaxation of the ESR state's population can thus be written as
\begin{equation}
\Gamma_\pm (\omega^{QI}_{\pm})  = \dfrac{\gamma^2}{2} \int dt \, e^{i \omega^{QI}_{\pm} t} \langle H^{QI}_{\mp}(t) H^{QI}_{\pm} (0) \rangle.
\end{equation}
Evaluating the dynamical field components in terms of magnetization deviations,
\begin{equation}
\begin{split}
H^{QI}_{x} &= \bar{B}_x^x \delta m'_x + \bar{B}_x^y \delta m'_y \qquad \qquad H''_{NV,y} = \bar{B}_y^x \delta m'_x + \bar{B}_y^y \delta m'_y \\
\Rightarrow \qquad \qquad & H^{QI}_{\pm} = [\bar{B}_x^x \pm i \bar{B}_y^x] \delta m'_x + [\bar{B}_x^y \pm i \bar{B}_y^y] \delta m'_y = \bar{B}_\pm^x \delta m'_x + \bar{B}_\pm^y \delta m'_y. 
\end{split}
\end{equation}
Therefore,
\begin{equation}
\begin{split}
\Gamma_\pm (\omega^{QI}_{\pm})  &= \dfrac{\gamma^2}{2} \int dt \, e^{i \omega^{QI}_{\pm} t} \langle H^{QI}_{\mp}(t) H^{QI}_{\pm} (0) \rangle \\
&= \dfrac{\gamma^2}{2} \int dt \, e^{i \omega^{QI}_{\pm} t} \langle [\bar{B}_\mp^x \delta m'_x(t) + \bar{B}_\mp^y \delta m'_y(t)] [\bar{B}_\pm^x \delta m'_x (0) + \bar{B}_\pm^y \delta m'_y(0)] \rangle \\
&= \dfrac{\gamma^2}{2} \int dt \, e^{i \omega^{QI}_{\pm} t} \left[ \bar{B}_\mp^x \bar{B}_\pm^x C'_{xx}(t) + \bar{B}_\mp^x \bar{B}_\pm^y C'_{xy}(t) + \bar{B}_\mp^y \bar{B}_\pm^x C'_{yx}(t) + \bar{B}_\mp^y \bar{B}_\pm^y C'_{yy}(t) \right]\\
&= \dfrac{\gamma^2}{2} \left[ \bar{B}_\mp^x \bar{B}_\pm^x C'_{xx}(\omega^{QI}_{\pm}) + \bar{B}_\mp^x \bar{B}_\pm^y C'_{xy}(\omega^{QI}_{\pm}) + \bar{B}_\mp^y \bar{B}_\pm^x C'_{yx}(\omega^{QI}_{\pm}) + \bar{B}_\mp^y \bar{B}_\pm^y C'_{yy}(\omega^{QI}_{\pm}) \right]\\
&= \dfrac{\gamma^2}{2} \sum_{i,j \in \{x,y\}} \bar{B}_\mp^i \bar{B}_\pm^j C'_{ij}(\omega^{QI}_{\pm}).
\end{split}
\end{equation}

\section{Parametrically driven in-plane easy-axis nanomagnet}
\textit{Eigenmode}| For the case of an in-plane easy-axis (z-axis) nanomagnet where the external field is applied along the z-axis ($\theta_H =0 $), we have the magnetization equilibrium oriented at angle $\theta_0=0$. In this case, the magnet frame (primed variables) and the lab frame (un-primed variables) are the same. The FMR frequency for this special case is given by 
\begin{equation}
\omega_\mrm{FMR} = \gamma \sqrt{\left[ H_0 + H_k \right]\left[ H_0 + H_k + H_\perp\right]}
\end{equation}
and the associated eigenvector is
\begin{equation}
\delta\vec{m}_0 = \dfrac{1}{\sqrt{\omega_\mrm{FMR}^2 + \gamma^2 H_{z,\mrm{eq}}^{2}}} \left( \begin{array}{c}
\gamma H_{z,\mrm{eq}}  \\ 
i \omega_\mrm{FMR} 
\end{array} \right) \equiv N \left( \begin{array}{c}
\omega_1  \\ 
i \omega_\mrm{FMR} 
\end{array} \right).
\end{equation}
The eigenvector denotes the mode profile of the FMR mode where 
\begin{equation}
-i\omega_\mrm{FMR} \delta\vec{m}_0 = \left( \begin{array}{cc}
0  & -\omega_1 \\ 
\omega_2 &  0 
\end{array} \right) \delta \vec{m}_0,
\end{equation}
where $\omega_1 = \gamma H_{z,\mrm{eq}} = \gamma ( H_0 + H_k ) $ and $\omega_2 = \gamma H_{z,\mrm{eq}} = \gamma ( H_0 + H_k + H_\perp) $, and $N =  1/\sqrt{\omega_\mrm{FMR}^2 + \gamma^2 H_{z,\mrm{eq}}^{2}}$. Based on the mode profile, we can define an ellipticity parameter $e = \vert \delta m_{0x}/ \delta m_{0y} \vert = \sqrt{\omega_1/\omega_2}$. Note that $e=1$ implies circular magnetization precession.\\

\textit{Magnetization deviation amplitude}| In this case, we can express the magnetization as
\begin{equation}
\begin{split}
\vec{m} (t) &= m_z \hat{z} + c(t) \, \delta\vec{m}_0 + c^*(t) \, \delta\vec{m}^*_0 \\
&\approx \left[ 1 - \vert c \vert^2 - \dfrac{\omega_1-\omega_2}{\omega_1+\omega_2} [c^2+c^{*2}] \right] \hat{z} + N \omega_1 [c(t)+c^*(t)] \hat{x} + i N \omega_\mrm{FMR} [c(t)-c^*(t)] \hat{y}.
\end{split}
\end{equation}
Note that $m_z(t)$ is not constant here. Since $\omega_1+\omega_2 \gg \vert \omega_1-\omega_2 \vert$, to leading order in nonlinearity 
\begin{equation}
\begin{split}
\vec{m} (t) &\approx \left[ 1 - \vert c \vert^2 \right] \hat{z} + N \omega_1 [c(t)+c^*(t)] \hat{x} + i N \omega_\mrm{FMR} [c(t)-c^*(t)] \hat{y}.
\end{split}
\end{equation}

Substituting the above expression in LL equation
\begin{equation}
\dfrac{d \vec{m}}{dt} = - \gamma \, \vec{m} \times \vec{H}_\mrm{eff}, 
\end{equation}
which in terms of Cartesian components is
\begin{equation}
\begin{split}
\dfrac{d m_x}{dt} &= - \gamma \, m_y (H_0 + H_k m_z)  \\
\dfrac{d m_y}{dt} &=  \gamma \, [ H_0 m_x  + (H_k + H_\perp(V)) m_z m_x ]. \\
\end{split}
\end{equation}
Note that $H_\perp(V) = H_\perp + \xi V$ where $V(t)$ is the applied voltage and $\xi = 2\beta/(M_s t_{ox}d_F)$ is the voltage induced change in effective perpendicular anisotropy field [$\beta$ is the VCMA coefficient, $t_{ox}$ is the oxide thickness, and $d_F$ is the ferromagnet thickness]. Hence,
\begin{equation}
\begin{split}
\dfrac{d m_x}{dt} &= - \omega_1 \, m_y - \gamma H_k [m_z-1] m_y \\
\dfrac{d m_y}{dt} &=  \omega_2\,  m_x + \gamma\,  \xi V m_x + \gamma [H_k + H_\perp + \xi V] [m_z-1] m_x.\\
\end{split}
\end{equation}

Substituting the ansatz for magnetization, 
\begin{equation}
\begin{split}
m_x &= N \omega_1 [c(t) + c^*(t)] \\
m_y &= iN \omega_\mrm{FMR} [c(t)-c^*(t)] \\
m_z &= 1 - \vert c \vert^2,
\end{split}
\end{equation}
we get
\begin{equation}
\begin{split}
\left[ \dfrac{d c}{dt} + \dfrac{d c^*}{dt} \right] &= - i \omega_\mrm{FMR} \, [c-c^*] +i \gamma H_k \dfrac{\omega_\mrm{FMR}}{\omega_1} \vert c \vert^2  [c-c^*]\\
\left[ \dfrac{d c}{dt} - \dfrac{d c^*}{dt} \right]  &=  - i \omega_\mrm{FMR} \,  [c+c^*] -  i \dfrac{\omega_1}{\omega_\mrm{FMR}} \gamma\,  \xi V [c+c^*] + i \gamma [H_k + H_\perp + \xi V] \dfrac{\omega_1}{\omega_\mrm{FMR}} \vert c \vert^2 [c+c^*].\\
\end{split}
\end{equation}
Keeping terms only to lowest order in nonlinearity [dropping terms $\sim V \vert c \vert^2 [c+c^*]$], we get
\begin{equation}
\begin{split}
\left[ \dfrac{d c}{dt} + \dfrac{d c^*}{dt} \right] &= - i \omega_\mrm{FMR} \, [c-c^*] +i \gamma H_k \dfrac{\omega_\mrm{FMR}}{\omega_1} \vert c \vert^2  [c-c^*]\\
\left[ \dfrac{d c}{dt} - \dfrac{d c^*}{dt} \right]  &=  - i \omega_\mrm{FMR} \,  [c+c^*] -  i \dfrac{\omega_1}{\omega_\mrm{FMR}} \gamma\,  \xi V [c+c^*] + i \gamma [H_k + H_\perp ] \dfrac{\omega_1}{\omega_\mrm{FMR}} \vert c \vert^2 [c+c^*].\\
\end{split}
\end{equation}

As a consequence,
\begin{equation}
\begin{split}
\dfrac{d c}{dt}  &= -i \omega_\mrm{FMR} \, c  -  i \dfrac{\omega_1}{2\omega_\mrm{FMR}} \gamma\,  \xi V [c+c^*] + i \left[ \gamma H_k \dfrac{\omega_\mrm{FMR}}{2\omega_1} + \gamma(H_k+H_\perp) \dfrac{\omega_1}{2\omega_\mrm{FMR}}\right] \vert c \vert^2 \, c \\
&+ i \left[ \gamma(H_k+H_\perp) \dfrac{\omega_1}{2\omega_\mrm{FMR}} - \gamma H_k \dfrac{\omega_\mrm{FMR}}{2\omega_1} \right] \vert c \vert^2 \, c^*, \\
\dfrac{d c^*}{dt}  &=  i \omega_\mrm{FMR} \,  c^* +  i \dfrac{\omega_1}{2\omega_\mrm{FMR}} \gamma\,  \xi V [c+c^*]  + i \left[ \gamma H_k \dfrac{\omega_\mrm{FMR}}{2\omega_1} - \gamma(H_k+H_\perp) \dfrac{\omega_1}{2\omega_\mrm{FMR}}\right] \vert c \vert^2 \, c \\
&- i \left[ \gamma(H_k+H_\perp) \dfrac{\omega_1}{2\omega_\mrm{FMR}} + \gamma H_k \dfrac{\omega_\mrm{FMR}}{2\omega_1} \right] \vert c \vert^2 \, c^* .\\
\end{split}
\end{equation}

The time-dependent voltage acts as a drive exciting magnons which we assume to be of the form $V(t)=V_0 \left[ e^{i\omega_p t} +  e^{-i\omega_p t}\right]/2$. Furthermore, to account for the fast oscillating time-dependence of the magnetization, we can define
\begin{equation}
c(t) = \tilde{c} e^{-i\omega_\mrm{FMR}t} \qquad c^*(t) = \tilde{c}^* e^{i\omega_\mrm{FMR}t}.
\end{equation}
In terms of $\tilde{c}$ and $\tilde{c}^*$,
\begin{equation}
\begin{split}
\dfrac{d \tilde{c} }{dt}   &=  -i \dfrac{\omega_1}{2\omega_\mrm{FMR}} \gamma\,  \xi \dfrac{V_0}{2} \left[ e^{i\omega_p t} +  e^{-i\omega_p t}\right]  \left[\tilde{c} + \tilde{c}^* e^{i2\omega_\mrm{FMR}t} \right] + i \left[ \gamma H_k \dfrac{\omega_\mrm{FMR}}{2\omega_1} + \gamma(H_k+H_\perp) \dfrac{\omega_1}{2\omega_\mrm{FMR}}\right] \vert \tilde{c} \vert^2 \, \tilde{c} \\
&+ i \left[ \gamma(H_k+H_\perp) \dfrac{\omega_1}{2\omega_\mrm{FMR}} - \gamma H_k \dfrac{\omega_\mrm{FMR}}{2\omega_1} \right] \vert \tilde{c} \vert^2 \, \tilde{c}^*  e^{i2\omega_\mrm{FMR}t} \\
\end{split}
\end{equation}
For parametric driving, $\omega_p \approx 2 \omega_\mrm{FMR}$ and keeping only slowly varying terms and dropping other fast oscillating terms,
\begin{equation}
\begin{split}
\dfrac{d \tilde{c} }{dt}   &=  -i \dfrac{\omega_1}{2\omega_\mrm{FMR}} \gamma\,  \xi \dfrac{V_0}{2} \, e^{-i\omega_p t}   e^{i2\omega_\mrm{FMR}t} \, \tilde{c}^*  + i \left[ \gamma H_k \dfrac{\omega_\mrm{FMR}}{2\omega_1} + \gamma(H_k+H_\perp) \dfrac{\omega_1}{2\omega_\mrm{FMR}}\right] \vert \tilde{c} \vert^2 \, \tilde{c}. \\
\end{split}
\end{equation}

In terms of amplitude $c(t)$ and $c^*(t)$,
\begin{equation}
\begin{split}
\dfrac{d c }{dt}   &= -i \omega_\mrm{FMR} \, c - i \dfrac{\omega_1}{2\omega_\mrm{FMR}} \gamma\,  \xi \dfrac{V_0}{2} \, e^{-i\omega_p t} \, c^*  + i \left[ \gamma H_k \dfrac{\omega_\mrm{FMR}}{2\omega_1} + \gamma(H_k+H_\perp) \dfrac{\omega_1}{2\omega_\mrm{FMR}}\right] \vert c \vert^2 \, c . \\
\end{split}
\end{equation}

Including damping of the magnon mode (with rate $\Gamma$) phenomenologically, we get
\begin{equation}
\begin{split}
\dfrac{d c }{dt} + i [\omega_\mrm{FMR} - \Psi \vert c \vert^2] \, c  + \Gamma \, c(t) &= - i \dfrac{\omega_1}{2\omega_\mrm{FMR}} \dfrac{\gamma\,  \xi V_0}{2} \, e^{-i\omega_p t} \, c^*,  \\
\end{split}
\end{equation}
where
\begin{equation}
\Psi = \left[ \gamma H_k \dfrac{\omega_\mrm{FMR}}{2\omega_1} + \gamma(H_k+H_\perp) \dfrac{\omega_1}{2\omega_\mrm{FMR}}\right].
\end{equation}

\section{Numerical details}
We numerically solve the LLG equation within the macrospin approximation
\begin{equation}
\dfrac{d\vec{m}}{dt} = -\gamma \, \vec{m} \times \left[\vec{H}_\mrm{ext} + H_k m_z \hat{z} - H_\perp(V) m_x \hat{x} \right] - \gamma \, \vec{m} \times \vec{h} + \alpha \, \vec{m} \times \dfrac{d\vec{m}}{dt}.
\end{equation}
Here, $H_\perp(V)$ is the voltage dependent perpendicular anisotropy where $V$ is the time-dependent voltage which for the case of parametric driving is $V = V_0 \sin\left(2\omega_\mrm{FMR} t\right)$ and for the case of large angle driving is $V(t) = V_0 \sum_{n=0}^{N}[\Theta(t-n(\tau_p+\tau_w))-\Theta(t-n(\tau_p+\tau_w)-\tau_p)]$ where $\Theta$ is the Heaviside step function, $N$ is the maximum number of pulses applied, $\tau_p$ and $\tau_w$ are the time duration's for which the voltage is turned on and off, respectively. Numerically solving the LLG equation provides us with the time-dependent magnetization that includes both coherent and incoherent dynamics. 

This magnetization is then translated to dipolar field $\vec{H}^\mrm{QI} = \mathcal{B} \vec{m}$ at the QI via the dipolar tensor ($\mathcal{B}$). We then solve for the QI effective TLS dynamics by solving the master equation for the QI density matrix $\rho$ using Qutip [Ref. S8]: $\dot{\rho} = -i [\mathcal{H}_{QI}^- +\mathcal{H}_{m-QI}^- ,\rho]$ where $\mathcal{H}^-_{QI} = (\omega_-^{QI}/2) [\mathcal{I} - \sigma_z]$ and $\mathcal{H}^-_{m-QI} = \gamma H^{QI}_z [\mathcal{I} - \sigma_z] + \gamma [H_+^{QI} \sigma_- + H_-^{QI} \sigma_+]/(2\sqrt{2})$. The time-dependent density matrix solution is then used to evaluate the QI spin state populations as a function of time. The incoherent field from the nanomagnet causes decoherence in the QI spin state population.

We repeat solving these coupled equations to average over different realization of the thermal noise field and thus obtain the ensemble averaged QI spin state population dynamics.\\

\begin{flushleft}
\textbf{\large Supplementary Material References}\\

[S1] A. Rustagi, I. Bertelli, T. van der Sar and P. Upadhyaya, Phys. Rev. B {\bf 102}, 220403 (R) (2020).

[S2] B. Flebus and Y. Tserkovnyak, Phys. Rev. Lett. {\bf 121}, 187204 (2018).

[S3] S. Chatterjee, J. F. Rodriguez-Nieva, and E. Demler, Phys. Rev. B {\bf 99}, 104425 (2019).

[S4] L. D. Landau, E. M. Lifshitz, and L. Pitaevskii, Statistical physics: theory of the condensed state, Vol. 9 (Butterworth-Heinemann, 1980).

[S5] T. L. Gilbert, IEEE transactions on magnetics {\bf 40}, 3443 (2004).

[S6] W. F. Brown, Phys. Rev. {\bf 130}, 1677 (1963). R. Kubo and N. Hashitsume, Progress of Theoretical Physics Supplement {\bf 46}, 210 (1970).

[S7] J. Norpoth, S. Dreyer, and C. Jooss, Journal of Physics D: Applied Physics {\bf 41}, 025001 (2007); R. Engel-Herbert and T. Hesjedal, Journal of Applied Physics {\bf 97}, 074504 (2005).

[S8] J.R. Johansson, P.D. Nation, Franco Nori, Comp. Phys. Comm. {\bf 183}, 1760–1772 (2012); J.R.Johansson, P.D. Nation, Franco Nori, Comp. Phys. Comm. {\bf 184}, 1234 (2013).
\end{flushleft}

\end{document}